\documentclass[aps,prd,
superscriptaddress,groupedaddress,nofootinbib]{revtex4}
\usepackage{amsmath,amssymb}
\usepackage{epsfig,graphicx}
\usepackage{amssymb}
\usepackage{epsfig,color}
\usepackage{pstricks,graphicx,epsfig,color,amssymb,amsmath,amscd}
\usepackage[]{graphicx}

\usepackage{makeidx}

\newcommand{\be}{\begin{eqnarray}}
\newcommand{\ee}{\end{eqnarray}}
\newcommand{\rar}{\rightarrow}

\usepackage[]{caption}
\def\-g{\sqrt{-g}}
\renewcommand\rho{\varrho}

\usepackage{multirow}

\begin{document}

\title{\bf Massive photons and electrically charged black holes}
\author{A.~D.~Dolgov}
\email{dolgov@fe.infn.it}
\affiliation{Novosibirsk State University \\
	Pirogova ul., 2, 630090 Novosibirsk, Russia}
\affiliation{ Institute of Theoretical and Experimental Physics \\
	Bol. Cheremushkinsaya ul., 25, 113259 Moscow, Russia}%
\author{K.~S.~Gudkova}
\email{k.s.gudkova@inp.nsk.su}
\affiliation{Budker Institute of Nuclear Physics\\
   Acad. Lavrentieva Pr., 11, 630090 Novosibirsk, Russia}

\date{\today}

\begin{abstract}
The characteristic time of disappearance of electric field in massive electrodynamics during
the capture of electric charge by black holes is calculated. It is shown that this time
does not depend upon the photon mass. The electric field at large distances disappears with
the speed of light.
\end{abstract}

\maketitle

\section{Introduction \label{s-intro}}

According to the conventional theory black holes (BH) have three kinds of externally observable classic parameters, see e.g. the book~\cite{FN-BH}. These parameters are namely the black hole mass, angular momentum, and electric charge. Electric field at large distance from an electrically charged BH with charge $Q$ has the usual Coulomb form:
\be
{\bf E_0} = \frac{Q}{r^2}  \frac{\bf r}{r}.
\label{E-m0}
\ee

However, this is true only if the photon mass is exactly zero. For massive photons, no matter how small is their mass,
electric field of a charged BH would vanish identically. Indeed, as it is shown in refs.~\cite{vilenkin,ll,dmt}, the finite energy solution of the Maxwell equations for the system of the Schwarzschild black hole and electrically charged thin spherical shell, concentric with the BH, results in the following expression for the electric field:
\be
{\bf E_m} = \frac{Q}{r^2}  \frac{\bf r}{r} \left( 1 - \frac{r_g}{R_s} \right) .
\label{E-m}
\ee
where $R_s $ is the radius of the shell, $r_g = 2M/m_{Pl}^2 $ is the BH gravitational radius, $M$ is the mass of the BH, and $m_{Pl} = 1.22\cdot 10^{19} $ GeV  is the Planck mass. This expression is valid at distance larger than gravitational radius $r\gg r_g$ but smaller than the inverse photon mass, $r m \ll 1$. For finite $m$ electric field at $r m \gtrsim 1$ turns into exponentially decreasing Yukawa field, $E \sim \exp (-m r )(1+ m r) / r^2$.

Evidently, when $r_s \rar r_g$, electric field disappears everywhere outside the BH. However, it is not evident how fast it disappears. It was conjectured that the characteristic time of disappearance is inversely proportional to a power of $\tau\sim 1/m$. So in this sense the discontinuity at $m = 0$ is smoothed down. Simple dimensional considerations lead to the conclusion that this is the only reasonable option. Moreover, it is claimed in ref.~\cite{pawl} that this is indeed the case. On the other hand, the Proca Lagrangian for massive vector field contains $m^2$ and no first power of $m$. So the next simplest hypothesis would be $\tau\sim 1/(m^2 d)$, where $d$ is some quantity with dimension of length which is related to the problem. However, according to the calculations presented in this paper electric field disappears with the speed of light, $\tau = r/c$, independently of $m$. Similar result is advocated in recent paper~\cite{erofeev}.

We will not discuss the origin of non-zero $m$ taking the simple case of hard switch-on the mass, simply by adding the mass term into the Maxwell Lagrangian. Gauge invariance in this case is evidently broken, but theory remains renormalizable. As we see, current conservation leads in this case to vanishing of the divergence of the potential $D_\mu A^\mu = 0$.

The paper is organized as follows. In. Sec. 2 we present massive field equations in the Schwarzschild metric. Sec. 3
includes solution for the Proca equation in case of a charged sphere collapsing to a black hole. In Conclusion we discuss possible physical consequences of this solution.

\section{Massive photons in curved space-time\label{eldyn-m-gamma}}

For massive photons the Maxwell equations turn into the Proca equation, which in curved space-time has the form:
\be
D_\mu F^{ \mu \nu} + m^2 A^\nu = 4\pi J^\nu
\label{D-F}
\ee
where $D_\mu$ is the covariant derivative in curved space-time, $F_{\mu\nu} = \partial_\mu A_\nu -  \partial_\nu A_\mu$,$A_\nu$ is the vector potential, and $J_\nu$ is the vector of  conserved electric current, $D_\nu J^\nu = 0$.

Current conservation demands the condition:
\be
\left( g^{\alpha \beta} D_\alpha D_\beta + m^2  \right)  D_\nu A^\nu  = D_\nu J^\nu = 0,
\label{d-J}
\ee
which is satisfied if $D_\nu  A^\nu = 0$.

We apply eq.~(\ref{D-F}) to calculation of electric field of spherically symmetric uniformly charged shell with initial radius $R_s$ and charge $Q$ with the Schwarzschild black hole in the center of the shell. We assume for simplicity that the mass of the charged shell is negligible in comparison with the mass of the BH. So the metric has the form:
\be
ds^2 = h(r)dt^2 - \frac{dr^2}{h(r) } - r^{2}d\Omega^2,
\label{metric}
\ee
where $h(r) = 1 - r_g/r$.

In spherically symmetric case the vector potential has only two nonzero components, $A_t (t,r) $ and $A_r (t,r) $, which satisfy the equations:
\be
h^{-1} \partial_t^2 E +m^2 \partial_t A_r &=&  \partial_t J_r \nonumber\\
\partial_r\left[h r^{-2} \left(E r^2 \right) \right] + m^2 \partial_r A_t &=& \partial_r J_t ,
\label{d-A-r-t}
\ee
where $E = \partial_t A_r - \partial_r A_t $ is the radial component of electric field and the current $J_\mu$ satisfies the conservation condition:
\be
\partial_t J_t - h r^{-2} \partial_r \left( r^2 h J_r \right) = 0.
\label{dt-Jt}
\ee

The electric field obeys the equation:
\be
\ddot E - h(r)\partial_r\left[{h(r)}{r^{-2}} \partial_r \left(r^2 E\right)\right] + h(r) m^2 E =
h \left( \partial_t J_r - \partial_r J_t \right).
\label{ddot-E}
\ee

\section{Electric field of a moving charged shell \label{s-shell}}

Thus we consider spherically symmetric uniformly charged shell with total charge $Q$. Inside this shell a  Schwarzschild black hole with mass $M$ and gravitational radius $r_g$ is located. Let us assume that initially at $t < 0$ the shell was at rest, then it started to move till $t = t_{max}$ and then stopped.
We do not specify the driving force of this motion and assume only that the mass of the charged sphere and the energy-momentum tensor of the source of the force are negligibly small.

The initial and final radii of the shell are respectively $R_{in} $ and $R_{fin}$. According to the solutions found in refs.~\cite{vilenkin,ll,dmt} the electric field at large distance is equal to
\be
 {\bf E_{in}} = \frac{Q}{r^2}  \frac{\bf r}{r} \left( 1 - \frac{r_g}{R_{in}} \right) .
\label{E-in}
\ee

After the shell stopped moving the solution at large time should tend to
\be
 {\bf E_{fin}} = \frac{Q}{r^2}  \frac{\bf r}{r} \left( 1 - \frac{r_g}{R_{fin}} \right) .
\label{E-fin}
\ee

We need to find how soon this asymptotic limit is reached.

In static case ($\dot E = 0$), far from the shell $r\gg R_c$ and with small $m r$  the solution takes the following form:
\be
E_0(r) = Q \left( 1 - \frac{r_g}{R_c} \right)\left( 1-r_gr m^2\right) \left( 1+m r \right) \frac{e^{-mr}}{r^2}
\approx \frac{Q}{r^2} \left( 1 - \frac{r_g}{R_c} \right) \left( 1 -\frac{m^2 r^2}{2} \right)   \left( 1- r_gr m^2\right).
\label{E-static}
\ee

If $R$ changes with time according to some not yet specified law, $R  = R(t)$, expression (\ref{E-static}) does not satisfy the full time-depending equation (\ref{ddot-E}) anymore. We look for the solution of this equation in the form:
\be
E(t,r) = E_{0} (t,r) + E_1 (t,r) .
\label{E-of-t-r}
\ee
where $E_{0} (t,r)$ is taken from the eq.~(\ref{E-static}) with time dependent $R = R(t)$ and hence
$E_1$ satisfies the equation:
\be
\ddot E_1 - h(r)\, \partial_r \left[\frac{h(r)}{r^2}\,\partial \left(r^2 E_1\right)\right] + h(r) m^2 E_1 = - \ddot E_0 (t,r).
\label{ddot-E1}
\ee
In what follows we take approximately
\be
E_{0} (t,r) =  \frac{Q}{r^2} \left( 1 - \frac{r_g}{R(t)} \right).
\label{E-stat-approx}
\ee

Under the specified conditions we expect that $E_1$ was zero at $t < 0$ and tends to zero for $t \rar +\infty$, so that the solution $E = E_{0} (t,r) + E_1(t,r) $ at large time tends to the stationary limit corresponding to $R=R_{fin}$.

We simplify eq.~(\ref{ddot-E1}) by introducing new function $Z(t,r)$ according to
\be
Z(r) = r \sqrt{h(r)} E_1,
\label{Z-of-r}
\ee
which satisfies the equation:
\be
\ddot Z - h^2 \left[ Z'' - Z\left(\frac{2}{r^2} - \frac{2 r_g}{r^3}\right)\right] + h m^2 Z = -  r \sqrt{h(r)} \ddot E_{stat} .
\label{ddot-Z}
\ee
Here prime means $d/dr$.

For large $r \gg R(t)$ and small $m$, such that $mr \ll 1$, we obtain:
\be
\ddot Z - Z'' + \frac{2 Z}{r^2} + m^2 Z = - r \ddot E_{0} (t,r).
\label{ddot-Z0}
\ee

We solve this equation perturbatively taking $Z = Z_0 + Z_1$, where $Z_0$ satisfies
\be
\ddot Z_0 - Z_0'' = - r \ddot E_{0} (t,r) \equiv  S(t,r).
\label{ddot-Z0}
\ee

Making the Fourier transformation
\be
\tilde Z_0 (\omega, r) &=& \int_{-\infty}^{+\infty} dt e^{ i \omega t} Z_0 (t, r) ,
\label{Z0-Fourier}
\\
Z_0 (t, r) &=& \int_{-\infty}^{+\infty} \frac{d\omega}{2\pi} e^{-i \omega t} \tilde Z_0 (\omega, r) .,
\label{Z0-inverse-Fourier}
\ee
we come to
\be
\tilde Z_0'' + \omega^2  \tilde Z_0 = -  \tilde S(\omega, r) ,
\label{tilde-Z0-dprime}
\ee
where $ S(\omega, r)$ is the Fourier transform of $S(t,r)$, defined in the r.h.s. of eq.~(\ref{ddot-Z0}).

This equation has the following solution:
\be
\tilde Z_0 = - \int_{r_0}^r dr_1 \frac{\tilde S(\omega, r_1) }{\omega} \sin \left[ \omega \left(r - r_1\right)\right] .
\label{Z-tilde-sol}
\ee

Now, returning to $Z_0 (t,r)$, we obtain:
\be
Z_0 (t,r) = Z_{hom}- \int_{-\infty}^{+\infty} \frac{d\omega}{2 \pi \omega} \int_{r_0}^r dr_1 \tilde S(\omega,r_1)\, e^{-i\omega t}\, \frac{e^{i \omega (r-r_1)} - e^{-i \omega (r-r_1)}}{2i},
\label{Z0-sol}
\ee
where
\be
Z_{hom} = f_1 (t-r) + f_2 (t+r)
\label{Z-hom}
\ee
is a solution of the homogeneous equation with arbitrary functions $f_1$ and $f_2$ which are to be
fixed by initial or final conditions. Evidently in massive electrodynamics longitudinal waves can
propagate over vacuum.

The contour of the integration over $\omega$ goes around the pole at $\omega = 0$ at positive $\rm{Im}\, \omega$. Such a choice is dictated by the condition of vanishing of $Z_{0}(t, r)$ at negative $t$. Indeed, the integration of the first exponent  in eq.~(\ref{Z0-sol}),  $\exp [i \omega (r-r_1)]$ gives  $-2\pi i \theta (t-t_1 - r + r_1)$, where $\theta (x) = 1$ if $x > 0$ and $\theta (x) = 0$ if $x<0$. So if $t< 0$, the unit step function vanishes and the contribution of the first exponent vanishes as well.

The second exponent in eq.~(\ref{Z0-sol}), $\exp [i \omega (r-r_1)]$ gives $-2\pi i \theta (t-t_1 + r - r_1)$ and since $r - r_1 >0$, this unit step function may be non-zero. However, we can write the integral over $dr_1$ as
\be
\int_{r_0}^r dr_1 = \left(\int_{r_0}^{+\infty} dr_1 - \int_{r}^{+\infty} dr_1\right) \theta(t+r - t_1  - r_1)...
\label{int-dr1}
\ee
The first term in the r.h.s. of eq.~(\ref{int-dr1}) is a function of $t+r$ and can be annihilated by a proper choice of $f_2 (t+r)$ and so we are left with the last integral for which $r_1 > r$. Hence the unit step function vanishes for $t<0$ as demanded.

It is convenient to rewrite the result for $Z_0 (t,r) $ as
\be
Z_0 (t,r) = -\frac{1}{2} \int_0^{t_m} dt_1 \left[ \int_{r_0}^r dr_1\, \theta (\Delta t - \Delta r) S(t_1,r_1) + \int_{r}^{+\infty} dr_1\, \theta (\Delta t + \Delta r) S(t_1,r_1) \right].
\label{Z0-1}
\ee

The lower limit of the integration over $dt_1$ in eq.~(\ref{Z0-1}) is taken equal zero because $S(t_1 <0) = 0$. Here, according to eqs.~(\ref{E-stat-approx}) and (\ref{ddot-Z0})
\be
S(t,r) = - \frac{Q r_g}{r} \frac{d^2}{dt^2} \left[\frac{1}{R(t)}\right],
\label{S-2}
\ee
and $\Delta t = t - t_1$, $\Delta r = r - r_1$. Note that in the first term of eq.~(\ref{Z0-1})
$\Delta r > 0$, while in the second term $\Delta r <0$.

Let us first take the integral over $dt_1$. In the first integral $\theta (\Delta t - \Delta r)$ ensures that $0 \leq t_1  \leq t -\Delta r$. We make a natural assumption that that the velocity $\dot R $ is continuous, so that  $\dot R(0) = \dot R(t_m) = 0$. Hence the integral is determined by the value of the integrand on the upper limit $t_1 = t -\Delta r$. The second integral is obtained from the first one by the interchange $\Delta r \rar -\Delta r$. So we find:
\be
Z_0 (t,r) =  \frac{Q r_g}{2}\left\{
\,\int_{r_0}^{r} \frac{d r_1}{r_1}\,\frac{d}{dt}\left[\frac{1}{R(t-r+r_1)}\right] +
\int_{r}^{+\infty} \frac{d r_1}{r_1}\,\frac{d}{dt}\left[\frac{1}{R(t + r-r_1)}\right] \right\} .
\label{Z0-2}
\ee
Evidently at $t<0$ both terms above vanish because of negative argument $\dot R(t) = 0$. Note that the
dependence on $r_0$ is spurious because it can be always moved to a function $f(t-r)$.

Let us now turn to positive $t$. The limits of the integration over $dr_1$ look simpler, if we rewrite the integrals changing variables as $r_1 \rar \rho = r -r_1$ and $r_1 \rar \rho = r_1 -r$ in the first and the second integrals respectively, which leads to
\be
Z_0 (t,r) =  \frac{Q r_g}{2}\left\{\,\int_{0}^{r-r_0} \frac{d \rho}{r -\rho}\,\frac{d}{dt}\left[\frac{1}{R(t-\rho)}\right] + \int_{0}^{+\infty} \frac{d \rho}{r + \rho}\,\frac{d}{dt}\left[\frac{1}{R(t-\rho)}\right] \right\} .
\label{Z0-3}
\ee
The integration limits over $\rho$ are determined by the interplay of two conditions:
\be
0 <\rho < r - r_0, \,\,\, {\rm and} \,\,\, t-t_m < \rho <t
\label{limits-1}
\ee
for the first integral and by:
\be
0 <\rho < \infty, \,\,\, {\rm and} \,\,\, t-t_m < \rho <t
\label{limits-2}
\ee
for the second one.

For very large time, such that $t  > r - r_0$ the following result is true:
\be
Z_0 (t,r) =  \frac{Q r_g}{2}\left\{\,\int_{t-t_m}^{r-r_0} \frac{d \rho}{r -\rho}\,\frac{d}{dt}\left[\frac{1}{R(t-\rho)}\right] + \int_{t-t_m}^{t} \frac{d \rho}{r + \rho}\,\frac{d}{dt}\left[\frac{1}{R(t-\rho)}\right] \right\},
\label{Z0-4}
\ee
The first integral is non-zero only if $r > t-t_m + r_0$. It does not fall down at large time. However, since $E \sim Z_0/r$, the electric field drops as $1/t$. The second term is smaller and drops as $1/t$. So we have found the necessary solution disappearing both at $t<0$ and at large positive $t$. This solution is a wave pulse with the width $L= t_m$ propagating with the speed of light.

Now we need to estimate $Z_1$ which satisfies the equation:
\be
\ddot Z_1 - Z_1'' = - \frac{2Z_0}{r^2} - m^2 Z_0
\label{ddot-Z-1}
\ee
Its solution is similar to the solution of eq.~(\ref{ddot-Z0}), see e.g. eq.~(\ref{Z0-1}) with $2Z_0(t_1,r_1)/r_1^2$ instead of $S(t_1,r_1)$. The term proportional to $m^2$ is neglected because we consider the limit of $m \rar 0$. Since, as it shown above, $Z_0$ is non-zero at $r\sim t$, the correction, $Z_1$, drops down with time faster than $Z_0$.

To sum up, when charged shell falls on the black hole, electric field on large distance vanishes with the speed of light.

\section{Conclusion \label{s-concl}}

Effective nonconservation of electric charge due to its complete disappearance inside black holes in massive electrodynamics could lead to cosmological electric asymmetry. Discussion of manifestations of this asymmetry and a list of early references can be found in~\cite{AD-DP}. In particular, electric asymmetry may be generated by excessive capture of protons in comparison with electrons because of much higher proton mobility in the cosmic plasma in analogy with the black hole charging considered in ref.~\cite{CB-AD-AP}.

Large scale electric fields in the charged universe may accelerate cosmic rays and the generated currents
could create large scale cosmic magnetic fields. Moreover, they might even induce accelerated cosmological expansion. These problems will be studied elsewhere.

\section*{Acknowledgement}
The work  was supported by the RSF Grant 20-42-09010.


\begin{thebibliography}{99}

\bibitem{FN-BH} V.P. Frolov, I.D. Novikov, "Black Hole Physics: Basic Concepts and  New Developments.", Springer, 1997.
\bibitem{vilenkin} A. Vilenkin, Phys. Rev. {\bf D20}, 373 (1979)
\bibitem{ll} B. Leaute and B. Linet, Gen. Rel. and Grav. {\bf 17}, 783 (1985).
\bibitem{dmt} A.D. Dolgov, H. Maeda, T. Torii, WU-AP-154-02, Oct 2002, \href{http://arxiv.org/hep-ph/0210267}{hep-ph/0210267}.
\bibitem{pawl} A. Pawl, Phys.Rev. {\bf D70} 124005, (2004), \href{http://arxiv.org/hep-th/0411175}{hep-th/0411175}.
\bibitem{erofeev} A.L. Erofeev, On dynamic aspects of the Proca field screening by a black hole,
Eur. Phys. J.C 80 (2020) 6, 495 • e-Print: 1911.03172 [gr-qc].
\bibitem{AD-DP} A. Dolgov, D.N. Pelliccia, Photon mass and electrogenesis Phys.Lett.B 650 (2007) 97-102 • e-Print: hep-ph/0610421 [hep-ph]
\bibitem{CB-AD-AP} C. Bambi, A.D. Dolgov, A.A. Petrov, Black holes as antimatter factories, JCAP 09 (2009) 013 • e-Print: 0806.3440 [astro-ph].
\bibitem{AD-JS-magn} A. Dolgov, J. Silk, Electric charge asymmetry of the universe and magnetic field generation, Phys.Rev.D 47 (1993) 3144-3150.

\end{thebibliography}
\end{document}